\documentclass[11pt,journal,onecolumn]{IEEEtran}

\usepackage{amsmath,amsfonts}
\usepackage{amsthm}
\usepackage{graphicx}
\usepackage{mathrsfs}
\usepackage{amssymb}
\usepackage{stfloats}
\usepackage[flushleft]{caption}
\usepackage[top=1in, bottom=1in, left=1in, right=1in]{geometry}
\theoremstyle{definition} 
\theoremstyle{definition} 
\theoremstyle{definition} 

\begin{document}

\title{A Message-Passing Approach to Combating\\
Hidden Terminals in Wireless Networks}
\maketitle

\begin{abstract}
Collisions with hidden terminals is a major cause of performance degradation in 802.11 and likewise wireless networks. Carrier sense multiple access with collision avoidance (CSMA/CA) is utilized to avoid collisions at the cost of spatial reuse. This report studies receiver design to mitigate interference from hidden terminals. A wireless channel model with correlated fading in time is assumed. A message-passing approach is proposed, in which a receiver can successfully receive and decode partially overlapping transmissions from two sources rather than treating undesired one as thermal noise. Numerical results of both coded and uncoded systems show the advantage of the receiver over conventional receivers.
\end{abstract}


\section{Introduction}
In 802.11 and likewise wireless networks, carrier sense multiple access (CSMA) is widely used to avoid collisions by requiring a node to verify the absence of concurrent transmission before transmitting on a shared transmission medium. However, not only does CSMA reduce spatial reuse, it also suffers from the hidden terminal problem, \textit{i.e.}, nodes may not sense each others' transmissions, even though they collide at the receiver. Methods such as raising the carrier sense threshold and RTS/CTS mechanisms are typically used to solve these problems. However, while we can enhance spatial reuse using the former method, it also causes more loss \cite{ref1}. In addition, experimental results show that enabling RTS/CTS significantly reduces the overall throughput \cite{ref2}\cite{ref3}, and hence WLAN deployments and access point (AP) manufacturers disable RTS-CTS by default \cite{ref4}.

We study techniques that allow a receiver to successfully receive and decode a desired packet in the presence of an overlapping packet from an interferer for a wide range of SNRs. Based on our previous work on joint channel estimation, interference mitigation and decoding \cite{ref5}, we propose a message-passing approach to combat collisions especially hidden terminals in wireless network. For wireless networks based on a protocol akin to current 802.11, our approach is able to recover both partially overlapping packets as long as the header of one packet is clear to the receiver under a wide range of SNRs. In practice, one header is often quite clear since it is located at the very beginning of the packet and its length is very short compared with that of the whole packet. Furthermore, if the protocol is designed to apply strong codes to protect the header, the same technique is likely to recover both packets even without any of
the headers being immediately decodable using single-user decoding.

Recent work \cite{ref6} proposes successive interference cancellation (SIC) to mitigate interference from other sources. SIC requires an explicit coordination between the interfering users, \textit{e.g.}, a significant difference in the power or coding levels \cite{ref7}, so that the packet transmitted with a higher power or lower rate can be decoded first and subtracted from the received signal before decoding the remaining packet or packets. However, in the scenario of hidden terminals, since the transmitters are hidden from each other, they cannot coordinate. When the two colliding packets are of comparable power levels or target rates, with high probability neither packet can be initially decoded through conventional SIC and thus no useful information can be retrieved.

Another way to combat hidden terminals is to use the ZigZag decoding scheme \cite{ref8}, which takes advantage of different interference-free chunks in retransmitted packets. However, retransmission reduces the overall throughput. Furthermore, in dealing with $M$ interferer, ZigZag decoder has to collect at least $M$ successive collisions of the same $M$ packets before any useful information can be retrieved, which is impractical for time-varying wireless channels.

The rest of this report is organized as follows. System model is formulated in Section II. Section III introduces specific design of receiver. Numerical results are provided in Section IV followed by conclusion in Section V.

\section{System Model}
In this section, we assume a system that is similar but different from 802.11 wireless network in certain ways. Consider the hidden terminal scenario in Fig. 1, where $T_{1}$ and $T_{2}$, unable to sense each other, transmit
simultaneously to the AP, causing collisions. Due to the lack of schedule in wireless network, AP receives the packet from $T_{1}$ partially overlapped with the packet from $T_{2}$. Assuming symbol synchronicity, the received signal at time $i$ in a frame length $l$ is expressed as
\begin{equation}
  \label{eq:sys}
  y_{i} = h_{i} x_{i} + h'_{i} x'_{i} + n_{i} \qquad i = 1 \ldots l
\end{equation}
where $x_{i}$ and $x'_{i}$ denote the transmitted symbols of user $T_{1}$ and user $T_{2}$, respectively, $h_{i}$ and $h'_{i}$ denote the corresponding $N_R$-dimensional vectors of channel coefficients whose covariance matrices are $\sigma_{h}^{2}\textbf{I} $ and $\sigma_{h'}^{2}\textbf{I} $, and $\{n_{i} \}$ represents the circularly-symmetric complex Gaussian~(CSCG) noise at the receiver with covariance matrix $\sigma_{n}^{2}\textbf{I}$. Since each received packet is subject to a delay, here $i$ denotes the absolute time. For simplicity, we assume an uncoded system with BPSK modulation, \emph{i.e.}, $x_{i}, x'_{i} $ are i.i.d. with values $\pm 1$ with respect to the overlapping part of the received data, while $x'_{i} $ is $0$ with respect to the interference-free part of $T_{1}$'s data and $x_{i} $ is $0$ with respect to the interference-free part of $T_{2}$'s data.

Assuming Rayleigh fading, $\{ h_{i} \}$ and $\{ h'_{i} \}$ are modeled as two independent Gauss-Markov processes, that is, they are generated by first-order auto-regressive relations ({\it e.g.}, \cite{ref11}):
\begin{subequations}
  \label{eq:auto-reg}
  \begin{align}
    h_{i}&= \alpha h_{i-1} + \sqrt{1-\alpha^{2}}\,w_{i}\\
    h'_{i}&=\alpha h'_{i-1} + \sqrt{1-\alpha^{2}}\,w'_{i}
  \end{align}
\end{subequations}
where $\{w_{i}\}$ and $\{w'_{i}\}$ are independent white CSCG processes with
covariance $\sigma_{h}^{2} \textbf{I}$ and $\sigma_{h'}^{2} \textbf{I}$, respectively, and $\alpha$
determines the correlation between successive fading coefficients. Typically, pilots are inserted periodically among data symbols. For example, 25\% pilots refers to pattern ``PDDDPDDDPDDD...'', where P and D mark pilot and data symbols, respectively.
\begin{figure}[!h]
\centering
\includegraphics[width=0.6\textwidth]{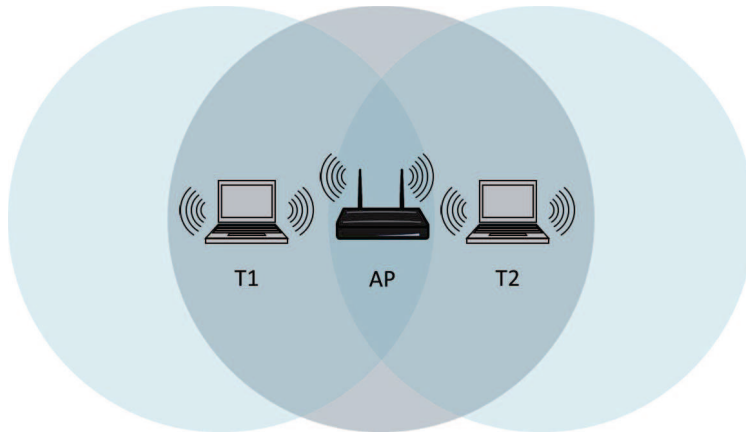}
\caption{A Hidden Terminals Scenario.}
\label{SYSTEMMODEL}
\end{figure}

\section{Receiver Design Based on Message Passing}
In this section, we will specifically introduce the issues in recovering both collided packets.
\subsection{Collision Detection}
To detect a collision, the receiver takes advantage of the known preamble of every packet \cite{ref10}. Here we use cross-correlation which is popular in detecting signal to detect collision due to the correlation property of the preamble. Suppose the length of preamble is $L$, AP aligns these L samples with the first $L$ samples of received data and computes the correlation. Then the alignment is shifted by one sample each time and the correlation is recomputed until the end of the received data. The preamble is a pseudo-random sequence that is independent of shifted versions of itself, as well as data from $T_{1}$ and $T_{2}$. Hence the correlation is near zero except when the preamble is perfectly aligned with the beginning of a packet. Fig. 2 shows the correlation as a function of the position in the received signal.

\begin{figure}[!h]
\centering
\includegraphics[width=0.6\textwidth]{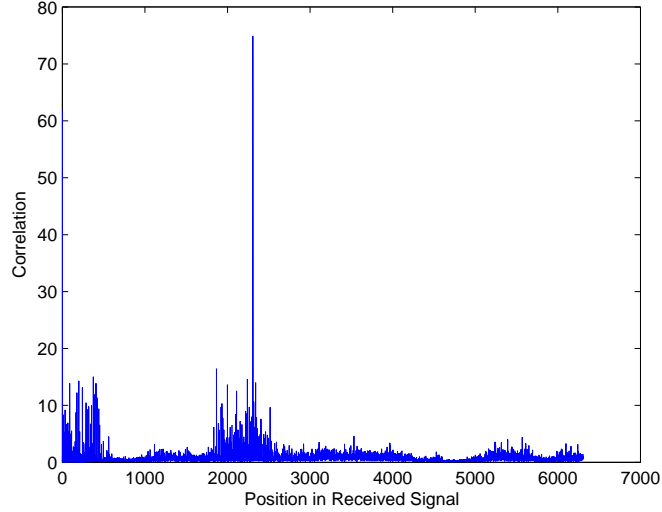}
\caption{Correlation function of position in received signal. The correlation spikes when the correlated preamble sequence aligns with the preamble in $T_{2}$'s packet, allowing the AP to detect the occurrence of a collision and where it starts.}
\label{SYSTEMMODEL}
\end{figure}

Note that when the correlation spikes in the middle of a reception, it indicates a collision. Hence,
the position of the spike corresponds to the beginning of the second packet. Mathematically, the correlation is computed as follows. Let $y$ be the received signal, which is the sum of the signal from $T_{1}$, $y_{1}$, the signal from $T_{2}$, $y_{2}$, and the noise term $n$. Let the samples $s[k]$, $1\leq k\leq L$, refer to the known preamble, and $s^{*}[k]$ be the complex conjugate. The correlation, $\Gamma$, at position $\Delta$ is:

\begin{equation}
  \label{eq:sys}
  \begin{aligned}
  \Gamma(\Delta)& = \sum_{k=1}^{L}s^{*}[k]y[k+\Delta]\\
                & = \sum_{k=1}^{L}s^{*}[k](y_{1}[k+\Delta]+y_{2}[k]+n[k])
  \end{aligned}
\end{equation}
Since the preamble is independent of the data of $T_{1}$ and the noise, the correlation between the preamble and these terms is close to zero. Since the first $L$ samples of $T_{2}$'s packet are the same as the preamble, we obtain:

\begin{equation}
  \label{eq:sys}
  \begin{aligned}
  \Gamma(\Delta)& = \sum_{k=1}^{L}s^{*}[k]y_{2}[k]\\
                & = \sum_{k=1}^{L}s^{*}[k]h'[k]s[k]\\
                & = \sum_{k=1}^{L}h'[k]|s[k]|^{2}
  \end{aligned}
\end{equation}
Since in practice, the channel gain does not vary very fast, the magnitude of $\Gamma(\Delta)$ is approximately the sum of energy in the preamble, and is relatively large, i.e., the magnitude of the correlation spikes when the preamble aligns with the beginning of $T_{2}$'s packet, as shown in Fig. 2. Imposing a threshold enables us to detect whether the AP received a collision signal and where exactly the second packet starts.
\subsection{Joint Recovery of Both Packets Using a Factor Graph}
Our approach is based on message passing and statistical inference on graphical models. This is a powerful framework that has found extensive use for decoding error correcting codes, and can incorporate a variety of $\textit{a priori}$ knowledge or side information about the interference, as well as the desired signal, along with unknown features that are estimated via message passing.

In prior work, we have applied this approach to the problem of detecting a narrow-band user in the presence of a narrow-band interferer, both of which experience dynamic fading \cite{ref5}. The conventional maximum likelihood detector is computationally impractical since it must search over (possibly a continuum) of combined channel and interference states. Instead, for this scenario, we can construct the factor graph shown in Fig. 3. The variables $h_{i}$, $x_{i}$, and $y_{i}$ are the channel and data for the first arrival packet, and the received signal, respectively, all at time i. The primes are associated with the latter arriving packet. The circular nodes represent the random variables shown, and the square nodes represent a probabilistic relationship among the variables connected to it \cite{ref9}. With each new observation, the conditional probability distributions shown in the figure can be updated according to the statistical assumptions concerning the channel and the noise. The distributions (or their simplified or parameterized approximations) are then passed to neighboring nodes, and
used to update the conditional distributions of the transmitted data.
\begin{figure}[!h]
\centering
\includegraphics[width=0.6\textwidth]{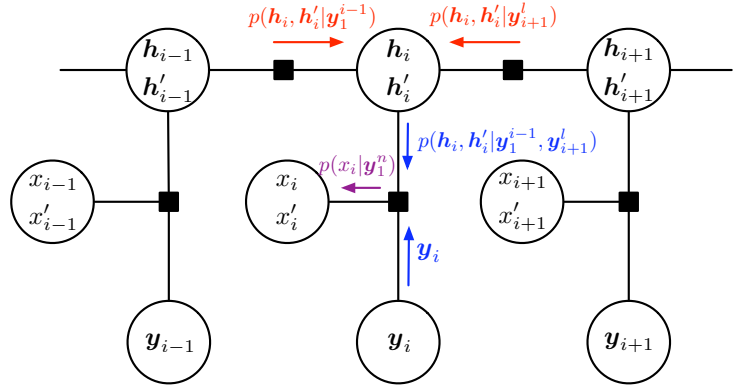}
\caption{Factor Graph Describing the Uncoded Communication System Model}
\label{SYSTEMMODEL}
\end{figure}
This message-passing algorithm, also referred to as belief propagation (BP), is equivalent to the sum-product algorithm widely used for decoding of error-correcting codes. Fig. 4 shows the factor graph incorporated with channel coding to jointly estimate and decode the received signal.

\begin{figure}[!h]
\centering
\includegraphics[width=0.6\textwidth]{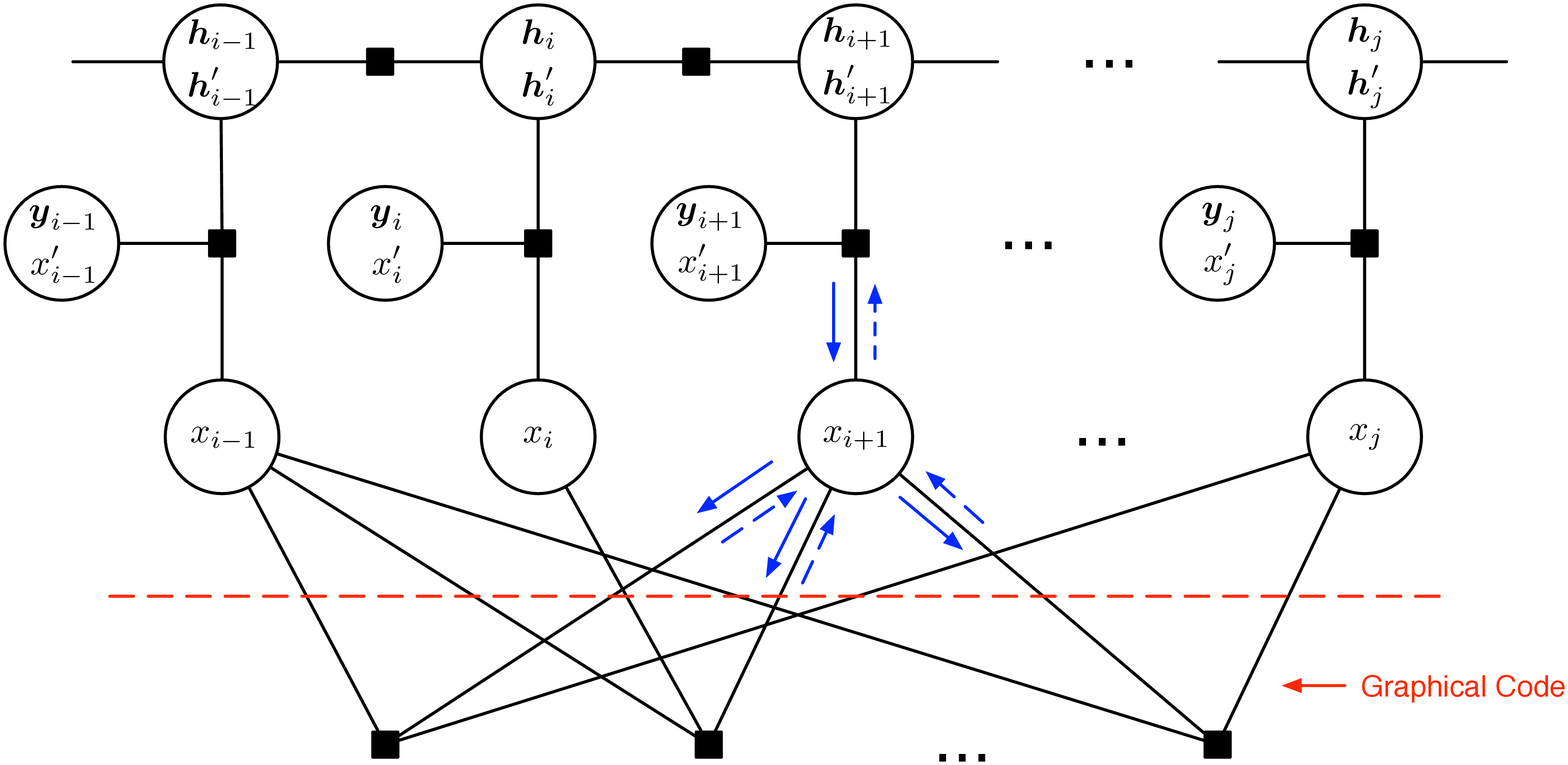}
\caption{Factor Graph Describing the Coded Communication System Model}
\label{SYSTEMMODEL}
\end{figure}

Due to the lack of synchronization in wireless LANs, the packets of $T_{1}$ and $T_{2}$ usually overlap with each other partially. The length of preamble and header is quite short compared with that of the whole packet. Therefore, we can easily decode the header of the first arrival packet and determine the length of that packet. Thus the exact segment of collision and the end of the first arrival packet are known. We can then initialize the prior probability of input signals ($x_{i}$ and $x'_{i}$) corresponding to different parts of the received data. By combining the factor graph and the message passing algorithm, we are able to estimate and decode both packets.

\section{Numerical Results}
In this section, the model presented in Section II and BPSK signaling, is used for simulation. We set the overlapping part of two packets to be uniformly distributed. The performance of the message-passing algorithm is plotted versus signal-to-noise ratio $SNR = \sigma_{h}^{2}/\sigma_{n}^{2}$, where the noise is i.i.d. We set the channel correlation parameter $\alpha = .99$ and limit the maximum number of Gaussian components to 8\footnote[1]{Because the messages are continuous probability density functions (PDFs), parametrization is used to characterize the PDFs exactly without introducing extra quantization error. Since the number of Gaussian components in the messages related to the fading coefficients grows exponentially. For implementation, we keep only a fixed number of components with the maximum amplitudes.}. Within each block, there
is one pilot in every 4 symbols. For the uncoded system, we set the frame length to $l = 4000$ with 56-bit preamble. For the coded system, we use a (500, 250) irregular LDPC code with 8-bit preamble and multiplex one LDPC codeword into a single frame.

\begin{figure}[!h]
  \centering
  \includegraphics[width=0.6\textwidth]{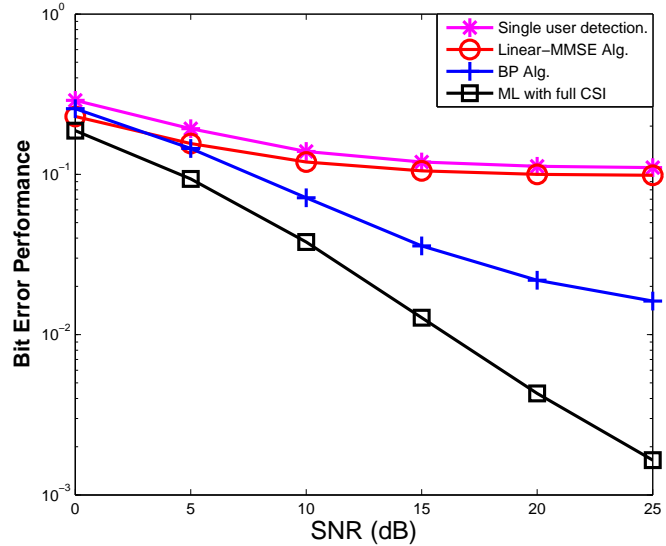}
  \caption{Bit error performance with an interferer, which is 3 dB weaker than the desired signal. The density of pilots is 25\%.}
  \label{fig:ch_m}
\end{figure}

\begin{figure}[!h]
  \centering
  \includegraphics[width=0.6\textwidth]{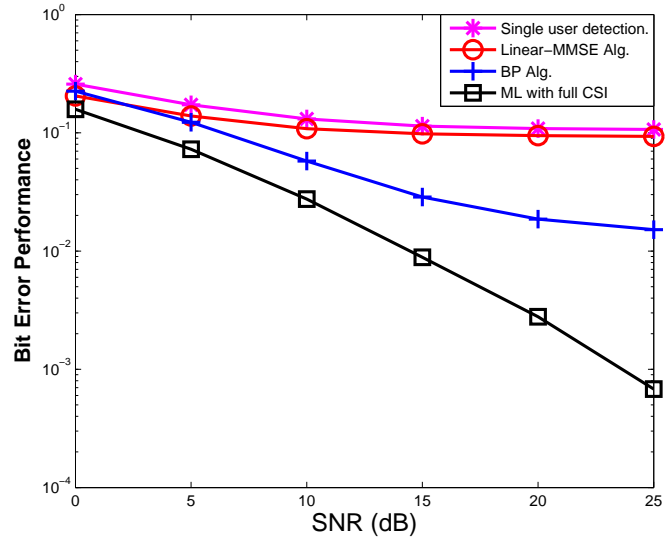}
  \caption{Bit error performance with an interferer, whose power is equal to the desired signal. The density of pilots is 25\%.}
  \label{fig:ch_m}
\end{figure}
\subsubsection{BER Performance}
Results for the message-passing algorithm with the Gaussian mixture messages described in Section III are shown in Figs. 5 to 8. We also show the performance of three other receivers for comparison. The first one is the conventional receiver which receives packets from one desired user and treats interference from undesired sources as noise. The second one is denoted by ``MMSE'', which estimates the desired channel by taking a linear combination of adjacent received value. This MMSE estimator treats the interference as white Gaussian noise. The third one is denoted by ``ML with full CSI'', which performs maximum likelihood detection for each symbol assuming that the realization of the fading processes is revealed to the detector by a genie, which lower bounds the performance of all other receivers.

Fig. 5 and Fig. 6 show uncoded BER vs. SNR, where the power of the latter arrival packet is $3$ dB weaker and equal to that of the first arrival packet, respectively. The message-passing algorithm generally gives a significant performance gain over the MMSE channel estimator and the conventional single user detection receiver, especially in the high SNR regime. Note that thermal noise dominates when the latter arrival packet is weak. Therefore, relatively little performance gain over the MMSE algorithm is observed in Fig. 5. In the very low SNR region, the MMSE algorithm slightly outperforms the message-passing algorithm, which is probably due to the limitation on number of Gaussian components. The trend of the numerical results shows that the message-passing algorithm effectively mitigates or partially cancels the interference at all SNRs of interest, as opposed to suppressing it by linear filtering.

\subsubsection{Channel Estimation Performance}
The channel estimate from the message-passing algorithm is much more accurate than that from the conventional linear channel estimations. Fig. 7 and Fig. 8 show the mean squared error for the channel estimation versus SNR where the interference signal is 3 dB weaker than and equal to the desired signal, respectively. One pilot is used after every three data symbols. Note that the performance of the linear estimators hardly improves as the SNR increases because the signal-to-interference-and-noise ratio is no better than 3 dB regardless of the SNR.  This is the underlying reason for the poor performance of the linear receiver shown in Fig. 5 and Fig. 6.
\begin{figure}[!h]
  \centering
   \includegraphics[width=0.6\textwidth]{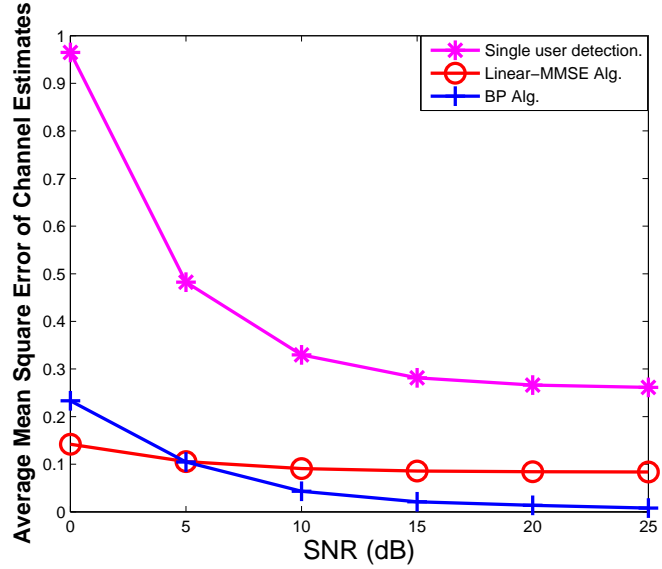}
  \caption{Channel estimation error with an interferer, whose power is 3dB weaker than the desired signal. The density of pilots is 25\%.}
   \label{fig:robust1}
\end{figure}

\begin{figure}[!h]
    \centering
    \includegraphics[width=0.6\textwidth]{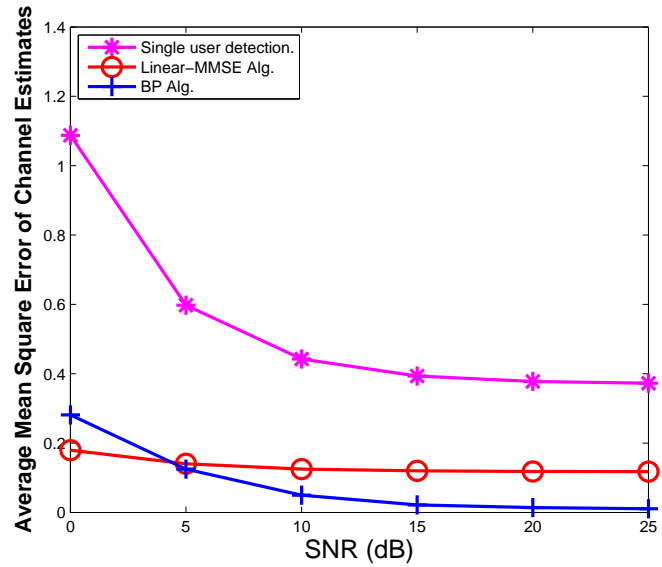}
    \caption{Channel estimation error with an interferer, whose power is equal to the desired signal. The density of pilots is 25\%.}
     \label{fig:robust2}
\end{figure}

\begin{figure}[!h]
    \centering
    \includegraphics[width=0.6\textwidth]{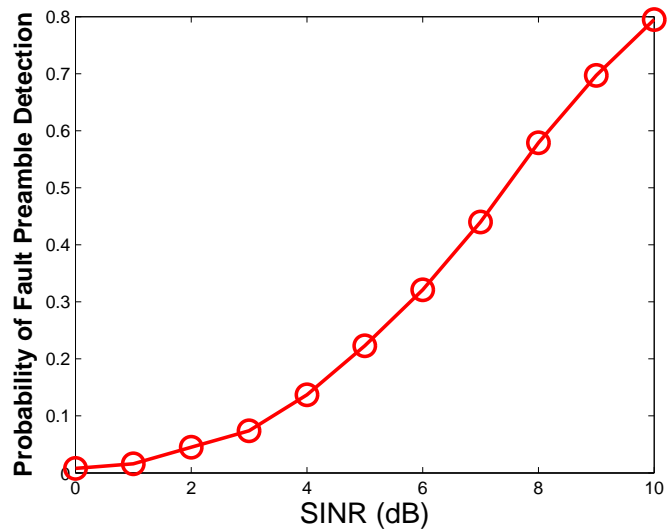}
    \caption{Probability of fault preamble detection vs SINR with signal to noise ratio 20dB.}
     \label{fig:robust2}
\end{figure}

\begin{figure}[!h]
    \centering
    \includegraphics[width=0.6\textwidth]{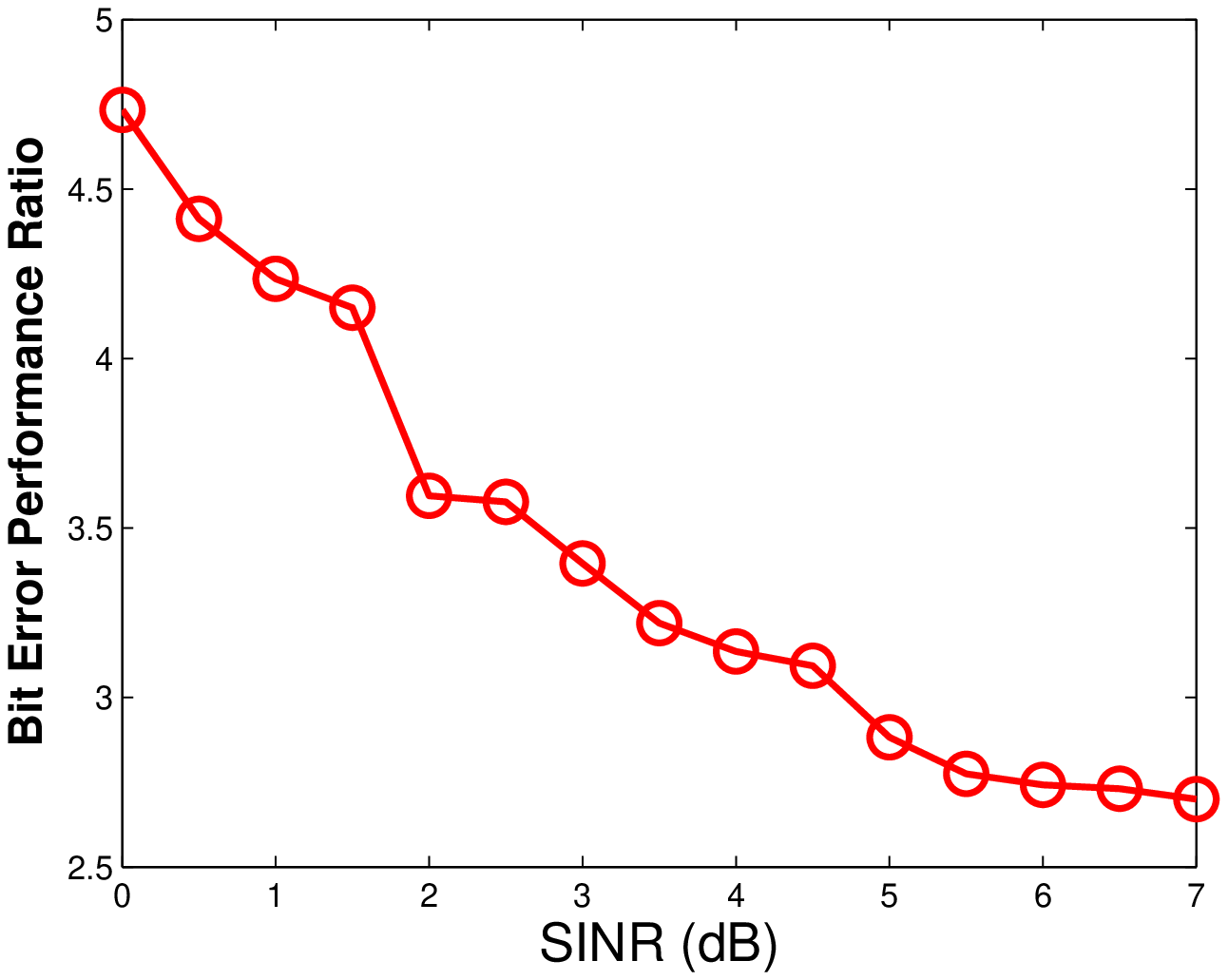}
    \caption{Bit error ratio of MMSE over BP with signal to noise ratio 20dB.}
     \label{fig:robust2}
\end{figure}

\subsubsection{Threshold of SINR}
When the interference signal is quite weak compared to the desired signal, we may not be able detect the start of collision. It is nearly optimal to treat weak interference as noise. Fig. 9 shows that when SINR goes to more than 5dB, the probability of fault preamble detection increases dramatically. It is useful to determine the threshold of SINR to indicate when collision detection and active mitigation is beneficial. We deem it meaningful when the bit error performance of our message passing algorithm is more than three times that of MMSE algorithm. From Fig. 10, we can see that, when we fix the signal-to-noise ratio to 20dB, the detection of collision makes sense only when the interference signal is no weaker than 5dB below the desired signal.

\begin{figure}[!h]
  \centering
  \includegraphics[width=0.6\textwidth]{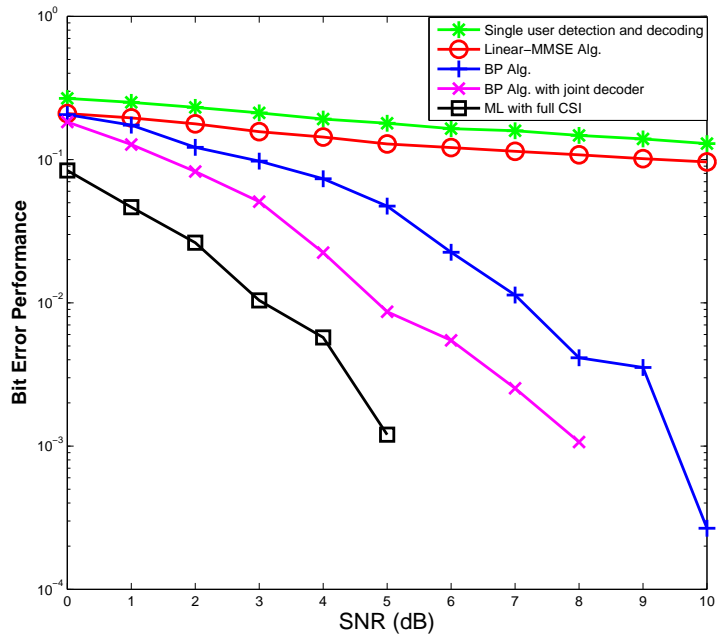}
  \caption{The BER performance for the system with a $(500, 250)$ irregular LDPC
    code. The power of interference is equal to that of desired signal. The density
    of pilots is 25\%.}
  \label{fig:ldpc1}
\end{figure}
Consider coded transmission using a $(500,250)$ irregular LDPC code and with one LDPC codeword in each frame,
\emph{i.e.}, no coding across multiple frames. Since we insert one pilot after every $3$ symbols, the total frame length is $667$ symbols. For the message-passing algorithm, let $I_{det}$ denote the total number of extrinsic information (EI) exchanges between decoder and detector, and $I_{dec}$ denote the number of iterations of the LDPC decoder during each EI exchange. Different values for pair $(I_{det}, I_{dec})$
correspond to different message-passing schedules.

In Fig. 11, we present the performance of two message-passing schedules: (a) $I_{det}=1$ and $I_{dec}=30$ denoted by ``Separate Message-passing Alg.'', \emph{i.e.}, the receiver detects the symbol first, then passes the likelihood ratio to the LDPC decoder without any further EI exchanges (separate detection and decoding), and (b) $I_{det}=3$ and $I_{dec}=10$, denoted by ``Joint Message-passing Alg.'', \emph{i.e.}, there are three EI exchanges and the LDPC decoder iterates $10$ rounds in between each EI exchange. For the other three receiver algorithms, the total number of iterations of LDPC decoder are both~$30$. As shown in Fig. 11, the message-passing algorithm preserves a significant advantage over the traditional linear estimator algorithms and the joint message-passing algorithm gains even more.

\section{Conclusion}
We have proposed a message-passing approach, in which a receiver can successfully receive and decode partially overlapping transmissions from two sources rather than treating another one as thermal noise. Numerical results of both coded and uncoded systems show the advantage of the receiver over conventional receivers.



\begin{thebibliography}{1}

\bibitem{ref1}
T.-S. Kim, H. Lim, and J. C. Hou. ``Improving spatial reuse through tuning transmit power, carrier sense
threshold, and data rate in multihop wireless networks''. \emph{In MobiCom}, 2006.

\bibitem{ref2}
G. Judd and P. Steenkiste. ``Using Emulation to Understand and Improve Wireless Networks and
Applications''. \emph{In NSDI}, 2005.

\bibitem{ref3}
P. C. Ng, S. C. Liew, K. C. Sha, and W. T. To. ``Experimental Study of Hidden node Problem in
IEEE 802.11 Wireless Networks''. \emph{In Sigcomm Poster}, 2005.

\bibitem{ref4}
Broadcom Wireless LAN Adapter User Guide.


\bibitem{ref5}
Y. Zhu, D. Guo and M. Honig. ``A message-passing approach for joint channel estimation, interference mitigation, and decoding''. \emph{Wireless Communications, IEEE Transactions on} , vol. 8, no. 12, pp. 6008--6018, 2009.

\bibitem{ref6}
D. Halperin, J. Ammer, T. Anderson, and D. Wetherall. ``Interference Cancellation: Better Receivers
for a New Wireless MAC''. \emph{In Hotnets}, 2007.


\bibitem{ref7}
S. Verdu, ``Multiuser Detection''. Cambridge University Press, 1998.

\bibitem{ref8}
S. Gollakota and D. Katabi, ``ZigZag Decoding: Combating Hidden Terminals in Wireless Networks'', \emph{ACM SIGCOMM}, vol. 38, no. 4, pp. 159-170, Oct. 2008.

\bibitem{ref9}
H.-A. Loeliger, J. Dauwels, J. Hu, S. Korl, L. Ping, and F. R. Kschischang, ``The factor graph approach
to model-based signal processing'', \emph{Proceedings of the IEEE}, vol. 95, no. 6, pp. 1295-1322, Jun. 2007.

\bibitem{ref10}
I. . WG. ``Wireless lan medium access control (mac) and physical layer (phy) specifications''.
\emph{Standard Specification, IEEE}, 1999.

\bibitem{ref11}
I. Abou-Faycal, M. M¡äedard, and U. Madhow, ¡°Binary adaptive coded pilot symbol assisted modulation over Rayleigh fading channels without feedback'', \emph{IEEE Trans. Commun.}, vol. 53, no. 6, pp. 1036¨C1046, June
2005.








\end{thebibliography}
\end{document}